\documentclass[letterpaper]{article}
\usepackage{graphicx}
\usepackage{hyperref}
\hypersetup{
    colorlinks=blue,
    linkcolor=blue,
    filecolor=blue,      
    urlcolor=blue,
    }
\usepackage{geometry}
\usepackage{algorithm}
\usepackage{algpseudocode}

\usepackage{caption}
\usepackage{amsmath}

\usepackage[longnamesfirst,sort]{natbib}% Citation support using natbib.sty
\bibpunct[, ]{(}{)}{;}{a}{,}{,}% Citation support using natbib.sty
\title{Inclusive education via empathy propagation in schools of students with special education needs}

\author{Igor Lugo\thanks{Corresponding author: \href{igorlugo@crim.unam.mx}{igorlugo@crim.unam.mx}}\\Universidad Nacional Autonoma de Mexico (UNAM)\\Centro Regional de Investigaciones Multidisciplinarias (CRIM)\vspace{0.5cm}\\ Martha G. Alatriste-Contreras\\Universidad Nacional Autonoma de Mexico (UNAM)\\ Facultad de Economia \vspace{0.5cm}\\Brenda G. Coutiño-Vázquez\\Universidad Nacional Autonoma de Mexico (UNAM)\\Centro Regional de Investigaciones Multidisciplinarias (CRIM)}

\date{November 20, 2025}

\begin{document}

\maketitle

\begin{abstract}
This study presents a theoretical model for identifying emergent scenarios of inclusiveness related to student with special education needs (SEN). Based on variations of the Shelling model of segregation, we explored the propagation of thinking about others as equals (empathy) in students with and without $SEN$ in school environments. We use the complex systems approach for modeling possible scenarios of inclusiveness in which patterns of empathy between students emerge instead of the well-known behavior of segregation. Based on simple transitional rules, which are evaluated by a set of null models, we show the emergence of empathy between students in school environments.
Findings suggest that small variations in the incentive of students for being considered as $SEN$ generate the presence of inclusive patterns. In other situations, patterns of segregation are commonly presented.
\end{abstract}

\section{Introduction}
The process for generating and applying a successful strategy of inclusiveness in education is not trivial. Students with special education needs (SEN)---``...disadvantages in physical, behavioral, intellectual, emotional and social capacities'' \citep{UNESCO2012}---commonly face discrimination and segregation situations in different educational environments. Despite the existence of inclusive educational policies \citep{UNESCO1994, EASNIE2025}, these are wrongly applied by the educators. The main justifications of this circumstance are the lack of trained personnel and number of the educators \citep{WardPowell2025}. In addition, there is a public perception of inclusiveness in schools that suggests a minimum effort and a quota for including students with SEN. Together with these facts, the local reality of schools is that they promote non-inclusive attitudes showing self-perpetuating patterns of segregation \citep{Goranssonetal2020}. Unfortunately, these situations are constantly reproduced at international, regional, and local scales suggesting that students with different capacities are left a side, delaying their development. It is evident that there is an inconsistency between the educational policies and the local facts of schools regarding the support of students with SEN. Therefore, we aim to show how likely an emergent pattern of inclusivity happens when students influence others by holding the opinion of being $SEN$. That is, the empathy as a conscious awareness of another person’s feelings and its real practice for understanding and working with others is the only way to make a difference for reversing the process of segregation in schools.

Based on the complex systems approach, in particular the cellular automata (CA) \citep{Wolfram2002}, the Schelling model \citep{Schelling1969, Schelling1978} and models of cultural dissemination \citep{Axelrod1984, LataneandNowak1997}, we shall explore different scenarios for generating inclusive pattern formations in the opinion of students with and without $SEN$. 
Students can show the potential of being empathic when they share a similar opinion of being $SEN$. 
Those scenarios will be related to deterministic transitional rules and different grid structures for understanding social dynamics related to the opinion formation and empathy dissemination for reversing the process of segregation. In particular, we are interested in answering the following questions: What are the deterministic rules that generate stable patterns of inclusivity? and what is the proportion of students with and without SEN that generates such stable patterns? In answering these questions, we use interdisciplinary and systemic approaches in which not only CAs models are the simple approximation for studying social interactions, but also they provide a better understanding of the rich and complex behaviors behind local influences and their emergent large-scale patterns \citep{Wolfram2020}.
In addition, we use a conceptual model validation that identifies the significance of our deterministic outputs by comparing them to expected null models---patterns generated from randomized data \citep{Farine2017,GotelliandGraves2020}.
Then, based on the fact that the Shelling's model can use similar proportion of agents with different opinions and symmetrical rules based on the number of neighbors with similar opinions for generating highly segregated patterns \citep{RogersandMcKane2011}, we expect to find that small variations in the number of students who hold the opinion of being $SEN$ influence the propagation of thinking about others as equals. This situation can be understood as an incentive to cooperate that generates patterns of inclusiveness in education. In addition, we believe that the segregation patterns of student persist because the inclusive strategies are wrongly executed and misunderstood by schools and its educators. This behavior can be related to an incentive to defect that generates the common patterns of segregation associated with the Shelling's model. In this particular case, it describes a minimum effort and a quota for including students with $SEN$ in schools. Therefore, if students are aware of being empathetic with other students, whether for their own account or on behalf of the inclusive strategies of schools, parents or their local community, they can modify deep-rooted patterns of segregation into inclusion.

This paper has been divided into four parts. The first part deals with the literature review. The second part describe the data that we use for generating our CA. The third part presents our method and data analysis. The fourth part shows results. The fifth part presents the discussion and conclusion sections.

\subsection{Literature review}
For simplicity, we present this section based on two types of literatures related to the public policies for inclusive education and the scientific studies of inclusivity in schools. Both literatures complement each other, but their contributions are diverse and controversial duo to a significant gap between the policy and the practices for inclusive education.

The former is based on international agreements, such as the \cite{UNESCO1994} and \cite{EASNIE2025}. They have proposed practical and strategic actions for providing education for all children. In addition, the \citet{OECD2015} reported inclusive innovation initiatives for improving the welfare of excluded groups in public services such as the education. This document suggests a balance between traditional innovation policies and a more inclusive innovation approaches.
In the context of Mexico, our current country of residence, the Political Constitution of the United Mexican States has recently amended Article 3, which states that education provided by the state must be ``compulsory, universal, inclusive, public, free, and secular,'' \citep{INEGI2023}. Even thought, since 2011, Mexico has used the framework of the General Law for the Inclusion of People with Disabilities, inclusive education has recently taken center stage between 2018 and 2024 due to the promotion of the National Strategy for Inclusive Education. The work of \citet{Zhizhko2020} described current efforts to carry out inclusive strategies in education, for example the Regular Education Support Services Unit and the Special Education and Inclusive Education Unit. However, the work of \citet{Morga2016} pointed out the lack of real and positive results in the inclusive education. 
%Mexico recently amended Article 3 of the Constitución Polítcia de los Estados Unidos Mexicanos, which states that education provided by the State must be "compulsory, universal, inclusive, public, free, and secular," as is the General Law of Education (DOF, 2019: 1). While Mexico has had the Ley General para la Inclusión de las Personas con Discapacidad since 2011, inclusive education has recently taken center stage between 2018 and 2024, during which the National Strategy for Inclusive Education (ENEI) was promoted.
%In the context of Mexico, our current country of residence, there is ``\href{https://www.diputados.gob.mx/LeyesBiblio/pdf/LGIPD.pdf}{La Ley General para la Inclusión de las Personas con Discapacidad, artículo 12}'' and the ``Constitución Política de los Estados Unidos Mexicanos, artículo 3'' that state the right to education and its inclusion.
%In particular, the work of \citet{Morga2016} pointed out the lack of real and positive results in the inclusive education. 
Consequently, even though the existence of these well-known international frameworks for actions, there is still an unequal access to education and segregated environments for students with SEN. The process for generating and executing inclusive strategies in schools is not trivial. It needs the coordination between teacher, educators, parents, society, and the state.

On the other hand, a large amount of scientific studies have stated the fact that the inclusive education should reduce significantly the deeply rooted path of segregation. In this respect, we can mention the work of \citet{Goranssonetal2020}. They showed that even if there is a strong commitment among teachers to the idea of an inclusive education, they still promote segregation behaviors due to a conscious or subconscious processes of resistance. In addition, the work of \citet{VanderBijetal2016} suggested that one of the keys for reducing the segregation in schools is a better execution of existing SEN strategies for the teachers. Furthermore, the work of \citet{VorlicekandKollerova2024} showed how an inefficient preparation of teachers and schools with socioeconomically disadvantages can result in the promotion of segregation of students with SEN even if teachers try to address the SEN requirements. Therefore, these references pointed out a real gap between the policy and the practices for inclusiveness. In particular, it shows the importance of teachers for coordinating inclusive activities and delivering a real knowledge and action of empathy. 

In the same line as above, there is a large number of studies related to the application of CAs in social sciences for modeling social dynamics, 
but there is a lack of theoretical formulations related to simulate the propagation of empathy between students. In this regard, there are fourth pioneering contributions that can be used for studying the empathy propagation between students in learning environments: the work of John Conway \citep{Gardner1970}, \citet{Schelling1978}, \citet{Axelrod1984}, and \citet{Krugman1996}. 
The ``Game of Life'' of Conway \citep{Gardner1970} was a contribution from the mathematics field in which Conway simulated real-life processes, in particular ``the rise, fall and alternations of a society of living organisms...'' He used a CA for showing how simple rules can generate complex and lifelike patterns.
The model of \citet{Schelling1978} showed how a small degree of homophily---individuals who prefer to interact with those more similar to themselves--- is sufficient to generate large-scale patterns of segregation. On the other hand, the model of \citet{Axelrod1984} used the idea of homophily and cultural assimilation---an increasing number of interactions between individuals can cause people to become more similar---for analyzing convergence and divergence of ideas or opinions. Finally, the work of \citet{Krugman1996} used the ``racetrack model'' for simulating the formation of cities based on a geographical context. More recent studies, for example the work of \citet{LugoAlatristeContreras2024}, analyzed the collective behavior of crowd disasters based on the concept of the personal space modeling in a CA. 
Therefore, these models show the potential of the CA for studying different and complex phenomena and a real possibility for analyzing educational inclusivity based on the dissemination of empathy between students. 

Therefore, in this study we aim to use a CA for describing the formation of inclusive patterns in schools based on the notion of empathy. Slight variations in the rules associated with the social influence---similarities with neighbors---and the personal resistance of changing the perception of being $SEN$  
generate stable patterns of convergence in empathetic opinions. Such stable patterns can be related to a set of successful strategies of inclusiveness in education. An example of a successful strategy based on an empathy-based social and emotional learning (SEL) program at Ireland is the study of \citet{Silkeetal2024}. 
In particular, the conscious attention and participation of teachers in the coordination of activities and the child care while students attend college are key elements for supporting and encouraging any type of inclusive strategies. That is, unplanned and uncontrolled interactions between students can reinforce the segregation patterns---conflicts---meanwhile planned and controlled interactions can generate inclusive patterns---cooperation.

\section{Materials}
We aim to show a theoretical approximation in which we explore different patterns of inclusion related to the self-perception of empathic students in schools. Consequently, we use computationally intensive experiments based on the Shelling's model of segregation for understanding the propagation of empathy and its resulted inclusive patterns in schools based on students with and without SEN. Therefore, we aim to generate our experimental database and then to determine its statistical significance based on a set of null models.

In addition, we aim to use Python third party libraries for generating our model, analyzing and showing our results. Particularly, the following libraries: \href{https://matplotlib.org/}{https://matplotlib.org/} and \href{https://numpy.org/}{https://numpy.org/}. For transparency, openness, and reproducibility of science \citep{Noseketal2015}, we share our code for modeling, analyzing, and showing results. This code is available in the following project of the Open Science Framework (OSF): see the Data availability section.

\section{Methods}
\begin{figure}[!h]
	\centering
	\includegraphics[width=0.65\textwidth]{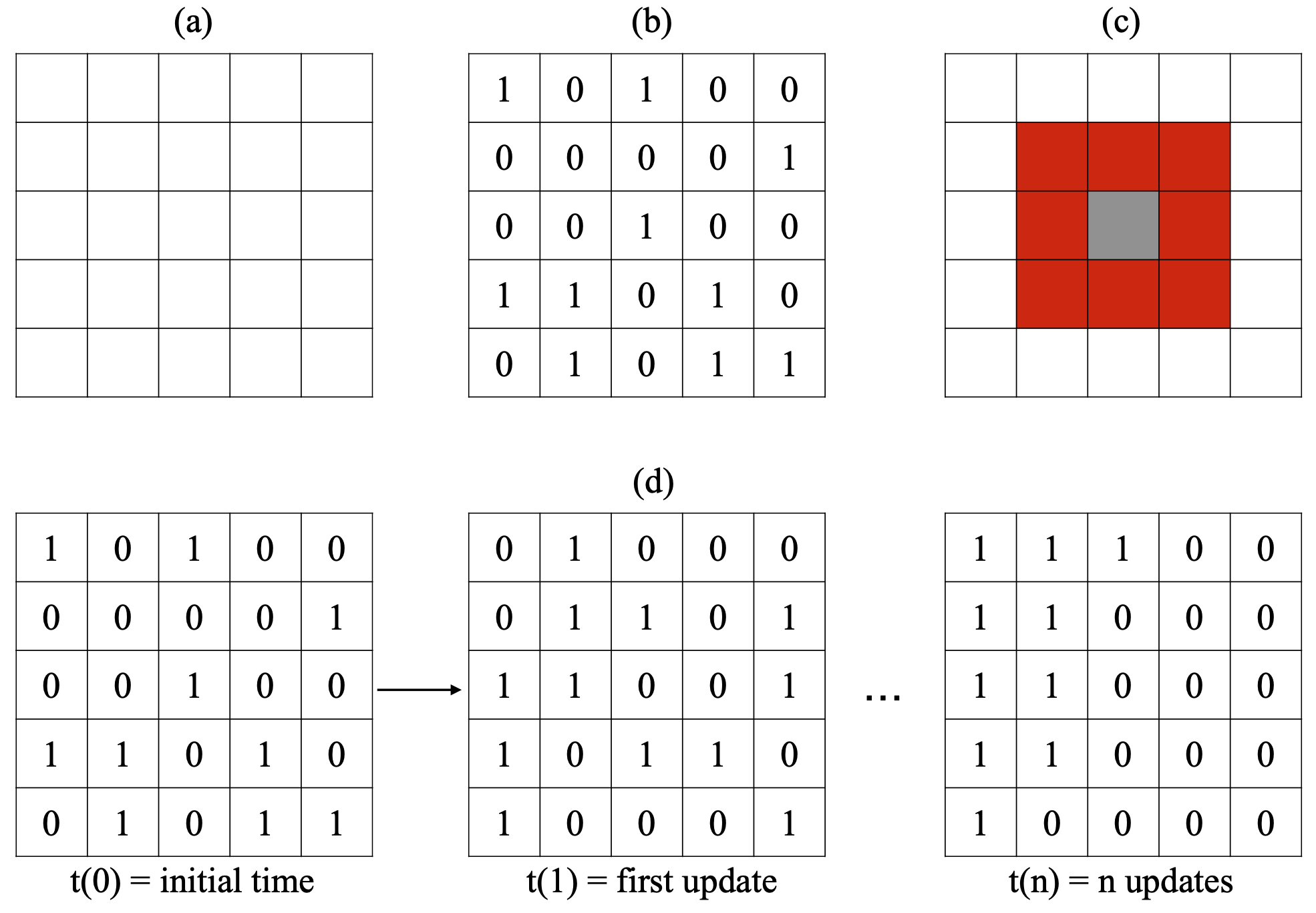}
	{\caption{An example of the basic features of CAs. (a) 2D array of 5x5 dimensions. (b) Each integer is associated with a qualitative data. (c) Local rules are applied to all cells. (d) Updating each array from $t(0)$ to $t(n)$.
	\label{cas}}}
\end{figure}

The current investigation uses CA for modeling the empathy propagation between students in school environments. These type of models are simple formulations and show high potential for understanding and describing social dynamics \citep{HegselmannFlache1998}. Their simplicity is based on the use of the following basic features (Figure \ref{cas}): 
\begin{enumerate}
	\item A grid that can show different types of dimensions and shapes (Figure \ref{cas} (a)).
	\item Each cell of this grid is associated with a state (integer or floating number) (Figure \ref{cas} (b)).
	\item States can be related to qualitative features associated with the analysis in question (Figure \ref{cas} (b)).
	\item States are updated (simultaneously or sequentially) base on local rules (Figure \ref{cas} (c) and (d)).
	\item Rules are applied to all cells (Figure \ref{cas} (c)).  
	\item Updating is related to the concept of ``time,'' which is discrete (Figure \ref{cas} (d)).
\end{enumerate}

Based on these basic features, we aim to generate a model for describing the propagation of empathy between students starting from the Shelling model of segregation. In the following subsection, we provide details about our theoretical model.

\subsection{Empathy propagation model}
Based on the Shelling's model of segregation \citep{Schelling1969, Schelling1978}, we aim to formulate the emergent process of inclusion based on the propagation of empathy between students with and without SEN in school environments. To describe this emergent process, we defined students (agents) based on two types of opinions (states) that they can hold, $SEN = 1$ or $nonSEN = 0$. Opinions can be related to a self-perception of being $SEN$ or $nonSEN$, and states are associated with two integer numbers representing the qualitative characteristic of the opinions. Then, we assume that an opinion can be changed by the presence and influence of neighbor opinions. This change of opinion, from $nonSEN$ to $SEN$, can be understood as the process of being empathetic. This process is commonly understood by a social mirroring---imitating others---and social cognition---sharing emotions with other \citep{Jankowiak2011}. In terms of the work of \citet{Axelrod1997} about cultural dissemination, there is a tendency for people to interact with similar ones, and if these interactions increase over time, there is a reinforcement of such similarities. 
On the other hand, the change from $SEN$ to $nonSEN$ also depends on the presence and influence of neighbor opinions, but it is more restrictive because when a student is truly emphatic with others is less likely to change to $nonSEN$. Based on the work of \citet{HortensiusdeGelder2018}, this process is related to the bystander effect: ``the reduction in helping behavior in the presence of other people''. 
Therefore, we propose a cell-like model in which the individual opinion of being perceived as $SEN$ or $nonSEN$ may change in the presence of a certain number of similar or dissimilar surrounding opinions.

Following the notation of \citet{Batty2005}, we begin with a grid (or array) in which each cell is associated with some state of a student, $SEN = 1$ or $nonSEN = 0$. These states are set randomly using a uniform random distribution in which these opinions are equally probable of the form $S_{ij}(0) = random(0,1)$, 
where $S_{i,j}(0)$ is an student located at (i,j) in time 0, and $random(0,1)$ is the state associated with the opinion. 

\begin{figure}[!h]
	\centering
	\includegraphics[width=0.55\textwidth]{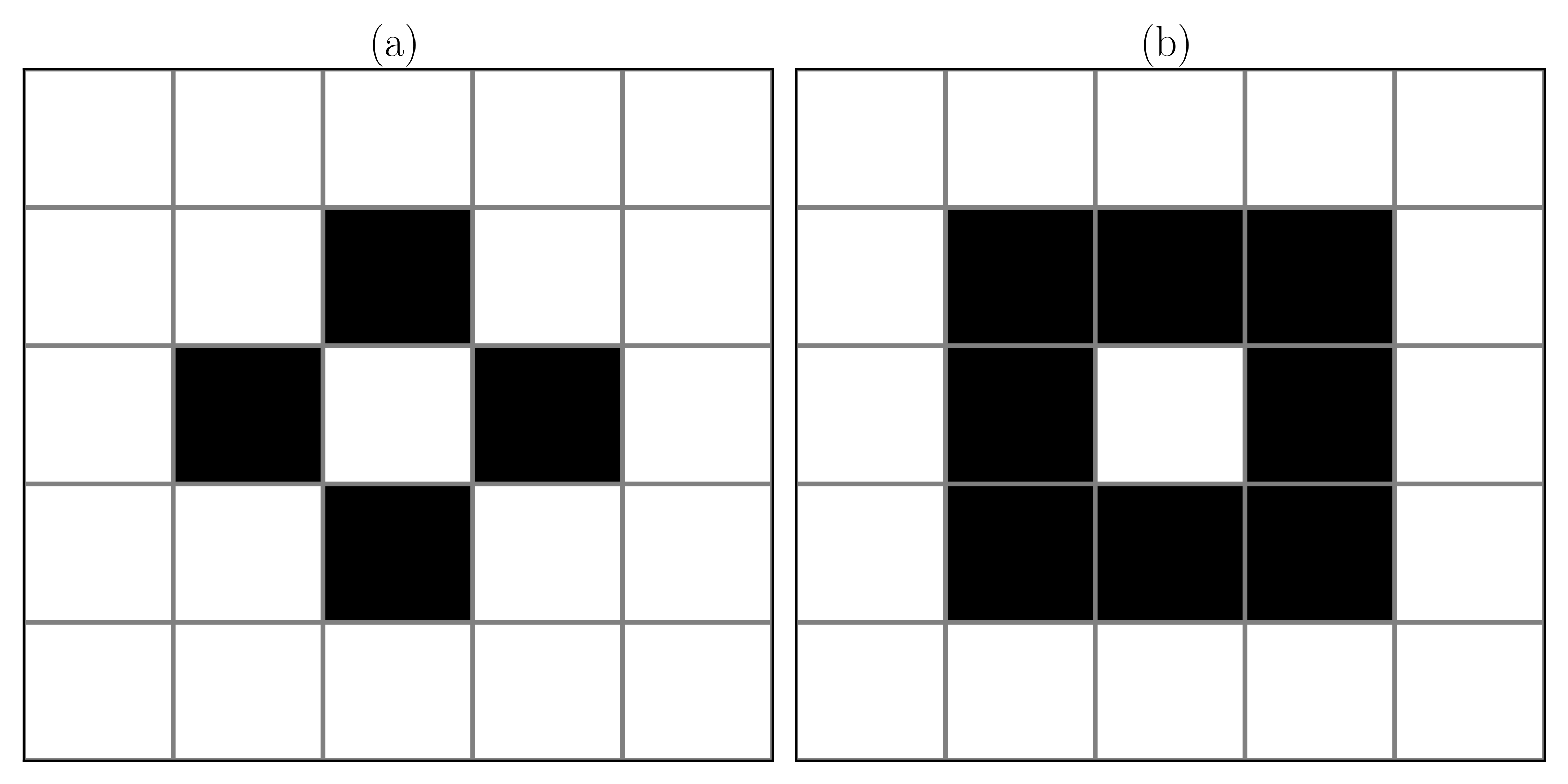}
	{\caption{Types of neighborhoods. (a) von Neumann. (b) Moore. In this case we are interested in using the Moore neighborhood.
	\label{typeneighbors}}}
\end{figure}

Next, a student at location (i,j) can change his/her opinion if a certain number of surrounding students in a Moore neighborhood shows different opinions(Figure~\ref{typeneighbors}). 
We selected the Moore neighborhood due to it can simulate the real-word dynamics of the spread of social behaviors, information, and cultural traits based on a local influence \citep{FlacheHegselmann2001}. 
Then, we defined this neighborhood based on the number of student neighbors related to the opinion of $SEN$ or $nonSEN$, $\bar{S}^{SEN}_{ij}(t)$ or $\bar{S}^{nonSEN}_{ij}(t)$, such as the following:

 \begin{equation} \label{eq:2}
 	\bar{S}^{SEN}_{ij}(t) = \sum_{k_{ij}\in \Omega \atop k_{ij} = SEN} S_{k_{ij}}(t) \quad \textrm{and } \quad \bar{S}^{nonSEN}_{ij}(t) = \sum_{k_{ij}\in \Omega \atop k_{ij} = nonSEN} S_{k_{ij}}(t)
 \end{equation}
where $k_{ij}$ is a location of an student opinion in the neighborhood $\Omega$, and $k_{ij}$ shows the opinion of $SEN$ or $nonSEN$.

Then, the change of his/her opinion is related to the number of surrounding students with similar opinion. Therefore, we updated each student opinion, $S_{ij}(t + 1)$, as follows:

\begin{equation} \label{eq:3}
S_{ij}(t + 1) = 
\left\{ 
  \begin{array}{ c l }
    0		& \quad \textrm{if } \;\; \quad S_{ij}(t) = 1 \quad \textrm{and } \quad \bar{S}^{SEN}_{ij}(t) < \textrm{$nonSEN_{students}$}\\
    		& \quad \textrm{elif } \quad S_{ij}(t) = 0 \quad \textrm{and } \quad \bar{S}^{nonSEN}_{ij}(t) = \textrm{$nonSEN_{students}$} \vspace{0.3cm}\\ 
    1		& \quad \textrm{if } \;\; \quad S_{ij}(t) = 0 \quad \textrm{and } \quad \bar{S}^{nonSEN}_{ij}(t) > \textrm{$SEN_{students}$}\\
    		& \quad \textrm{elif } \quad S_{ij}(t) = 1 \quad \textrm{and } \quad \bar{S}^{SEN}_{ij}(t) = \textrm{$SEN_{students}$}    
		\vspace{0.3cm}\\ 
  \end{array}
\right\}
\end{equation}
where $\bar{S}^{SEN}_{ij}(t)$ and $\bar{S}^{nonSEN}_{ij}(t)$ are the sum of the values of neighbors related to a student opinion, and the $SEN_{students}$ and $nonSEN_{students}$ are parameters for specifying different levels of opinions. In the Shelling's model of segregation, these parameters are commonly set by the number $4$. In our case, we propose to vary these parameters as follows.

%%%%
\subsection{From segregation to inclusion}
Based on the parameters $SEN_{students}$ and $nonSEN_{students}$, we can define fourth types of criteria related to a similar number of surrounding opinions, and each criteria described different scenarios of student behaviors, from segregation to inclusion. These scenarios are the following:

\begin{itemize}
	\item Escenario 1: Segregation. This scenario is the common result found in the Shelling model of segregation. In this scenario the criterion is set by the following parameters: $SEN_{students} = 4 $ and $nonSEN_{students} = 4$. This scenario describes the situation in which even when students are indifferent of being surrounded by different students, they choose to segregate themselves from other students. Therefore, students might self-segregated without a particular reason. 
	
	\item Escenario 2: Empathy and inclusivity based on a direct influence of $SEN$ students.
In this scenario the criterion is set by the following parameters: $SEN_{students} = 5 $ and $nonSEN_{students} = 4$. This scenario describes a situation in which the opinion of being $SEN$ is more accepted by students due to the influence of neighbors. In particular, the parameter $SEN_{students} = 5$ suggests higher peer pressure of $SEN$ students for accepting and changing opinions. Therefore, the empathy and inclusivity are generated based on a slightly higher influence from the $SEN$ students.
	
	\item Escenario 3: Empathy and inclusivity based on an indirect influence of $SEN$ students. This scenario is related to the following criterion associated with the following parameters: \\$SEN_{students} = 4$ and $nonSEN_{students} = 5$. This scenario describes a situation in which the opinion of being $SEN$ is less susceptible to change due to the benefits of being emphatic. In particular, the parameter $nonSEN_{students} = 5$ suggests higher restrictions for changing an opinion. Therefore, the empathy and inclusivity are generated based on a slightly higher restriction to be influenced from the $nonSEN$ students.
	
	\item Escenario 4: Empathy and inclusivity based on a direct and indirect influence of $SEN$ students. This scenario is the combination of the two above-mentioned scenarios. In this scenario, the parameters are a mix of the above criteria that are set by $SEN_{students} = 5 $ and $nonSEN_{students} = 5$. This scenario outlines a situation in which the opinion of being $SEN$ is more accepted and less susceptible to change by students. This scenario reinforce the process of thinking about others as equals. Therefore, the inclusivity is more visible and effective through an increased stimulus by the $SEN$ students.
\end{itemize}
 
In addition to these scenarios, we are interested in using different proportions of students with $SEN$ and $nonSEN$ opinions. Such proportions describe different initial conditions that can affect the number of iterations to reach a clear and stable scenario of empathy and inclusivity. In particular, we are interested in the following tuples that specify such proportions: (0.5, 0.5), (0.6, 0.4), and (0.4, 0.6). The first tuple shows a similar proportion between $SEN$ and $nonSEN$ opinions, and it is similar than the uniform random distribution that we mentioned in the subsection Empathy dissemination model; the second tuple displays higher proportion of $SEN$ than $nonSEN$; and the third tuple shows higher proportion of $nonSEN$ than $SEN$.

\subsection{Data analysis and model validation}
Based on the above criterion and scenarios, we apply an explorative data analysis and model validation \citep{Sargent1984, LeekandPeng2015}. In particular, the exploratory data analysis searches for trends and relationships between the common settings of the Shelling's model and the variations of these settings based on the different proportions of $SEN$ and $nonSEN$ opinions. To determine whether our theoretical model describes the presence of inclusivity, we should evaluate different values of parameters and 
a set of initial conditions for reaching a clear and stable scenario of inclusivity.

We use a conceptual model validation in which the theory and assumptions should generate consistent and justifiable results \citep{KerrGoethel2014}. For validating our model, we use three types of null models to determine if our results could have arisen by random chance. 
These models are related to the binomial distribution, which is the number of successes in a fixed number of independent Bernoulli trials, and different values of the parameter $p$---probability of success. The Bernoulli distribution is a discrete probability distribution in which a random variable $X$ can take only two values $X= 1$ for success and $X=0$ for failure. In particular, for the nature of our analysis---the presence of two possible outcomes, $SEN$ and $nonSEN$ students \citep{Maddenetal2019, Wallert2022}---we are interested in the probability of success when $p=0.3$, $p=0.5$, and $p=0.7$. The first case defines a rule in which there is 30\% probability of success for selecting $SEN$ students; the second case defines a rule in which there is 50\% probability of success or the same chance to select $nonSEN$ or $SEN$ students; and the third case describes a rule in which there is 70\% probability of success for selecting $SEN$ students. These cases explore randomized escenarios in which transitional rules are designed to generate biased or unbiased selection of $SEN$ and $nonSEN$ students. 
In this respect, we followed the main idea in \citet{Ross2019}, \cite{LugoMartinezMekler2022}, and \cite{LugoAlatristeContreras2022} who use different statistical distributions for identifying one of the most important distributions that can represent special cases and transformations from one distribution to another. Consequently, because of the use of different values of the parameter $p$---which determines the shape of the distribution, $p=0.5$ symmetric, $p<0.5$ positive skew, and $p>0.5$ negative skew \citep{MolugaramandRao}---the binomial distribution may provide the basic fundamentals for representing special cases and transformations of other discrete probability distributions. 

\begin{figure}[!h]
	\centering
	\includegraphics[width=0.18\textwidth]{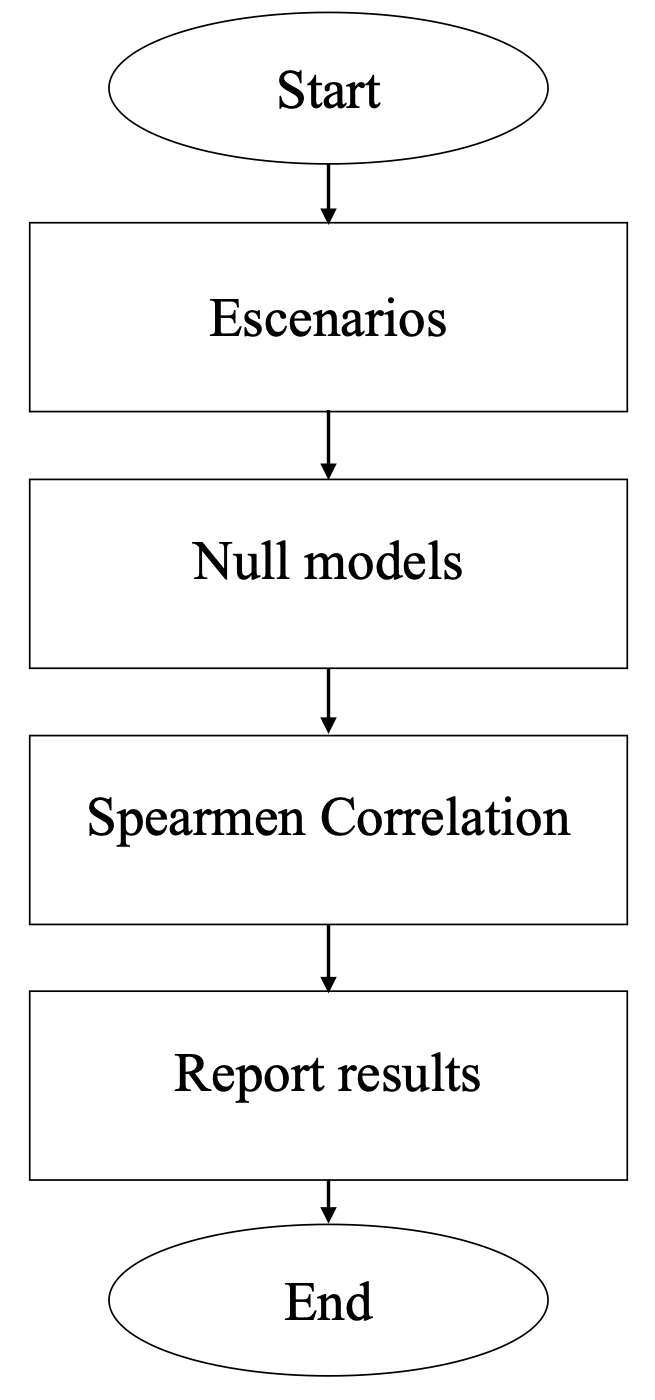}
	{\caption{Flowchart of the model validation.
	\label{mvalidation}}}
\end{figure}

Next, each of these null models uses a probabilistic framework based on a Monte Carlo approximation that looks for statistical regularities \citep{Gentle2022}. Comparing our results with our null models, we evaluate if our deterministic rules are different from patterns generated by random chance. In other words, we evaluate whether our transitional rules offer a significant explanation over a random description. 
%The presence of such regularities can validate the use of our initial conditions because the model outputs may agree with the observed data of public perception and policies of inclusivity. Therefore, our model validation process shall support our conclusions.
Furthermore, we use a sensitivity analysis that determines the relative influence of the initial conditions and the parameters on our model output. We are interested in exploring which combination of those parameters are related to a clear inclusive escenario. To do this, we shall follow the procedure outlined in Figure \ref{mvalidation}.  

The process starts comparing our generated escenarios with those null models. Each escenario is compared with every null model based on the Spearman correlation coefficient. We compute the median of $10,000$ realizations of each null model. Then, we measure the correlation coefficient between each escenario and every null model realizations. Next, the initial proportions of $SEN$ and $nonSEN$ students in each escenario should be varied to determine the relative influence of initial conditions. Finally, we report our results.

Another significant aspect of the data analysis and model validation is to set the initial conditions. For simplicity, the total number of cells is set by a matrix with fixed dimension equal to $100\times100$, that is $10,000$ cells. Because we are interested in the emergent process of thinking about others as equals in students with and without $SEN$ in school environments, we use a torus configuration---an array with no edges where left/right and top/bottom are connected. This topology represents an idealize environment where the student opinions can change without being limited by physical or social barriers \citep{FlacheHegselmann2001}. 
The number of iterations in each realization is fixed to a total of $15$ due to after this number of updates, the model converges into a singular pattern, a point of attraction.

\section{Results}
\begin{figure}[!h]
	\centering
	\includegraphics[width=0.60\textwidth]{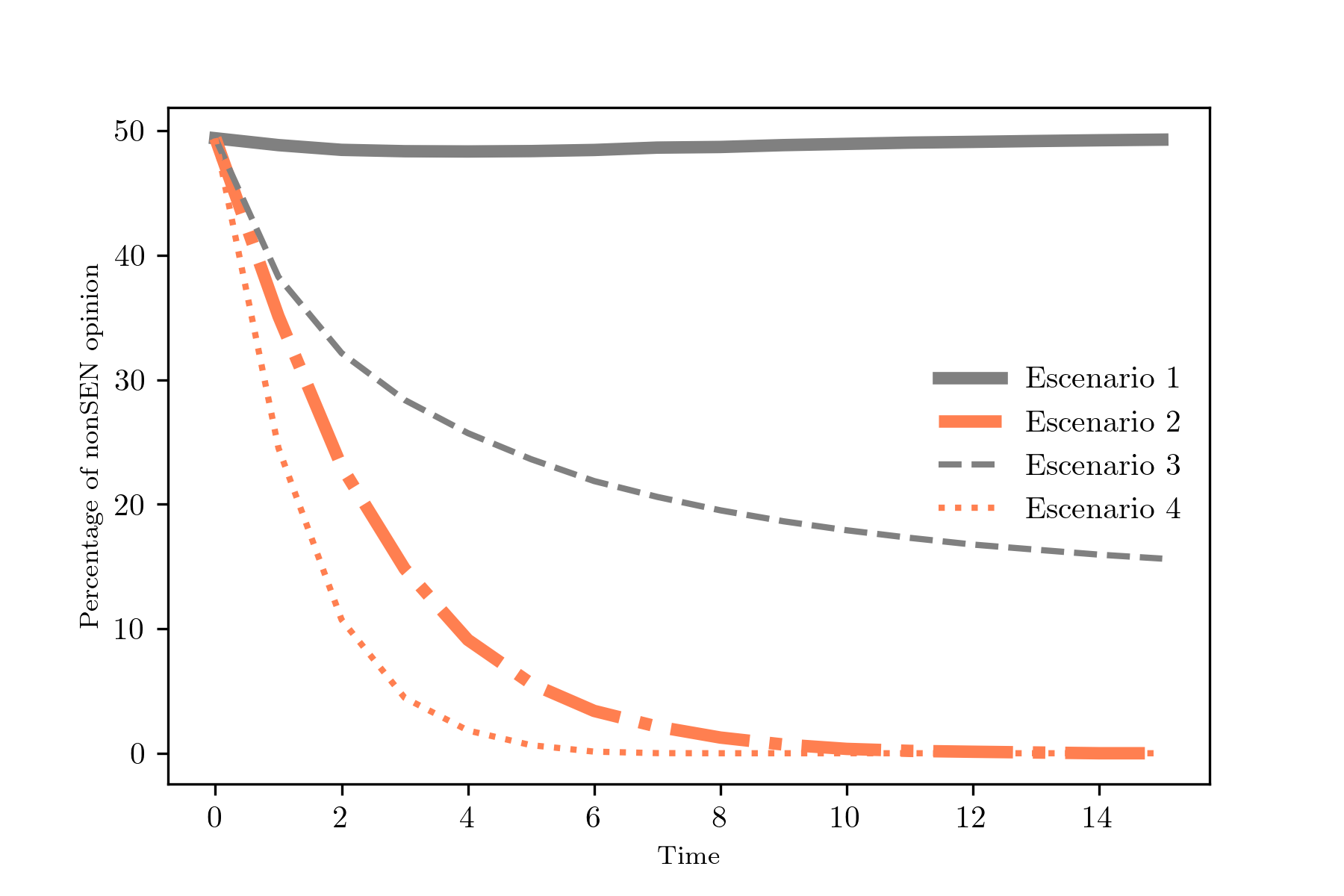}
	{\caption{Escenarios of empathy by percentage of $nonSEN$ students. Escenario 1, segregation: $SEN_{students} = 4 $ and $nonSEN_{students} = 4$. Escenario 2, convergence of empathy: $SEN_{students} = 5 $ and $nonSEN_{students} = 4$. Escenario 3, convergence towards empathy: $SEN_{students} = 4 $ and $nonSEN_{students} = 5$. Escenario 4, convergence of empathy: $SEN_{students} = 5 $ and $nonSEN_{students} = 5$.
	\label{es01}}}
\end{figure}

\begin{figure}[!h]
	\centering
	\includegraphics[width=0.70\textwidth]{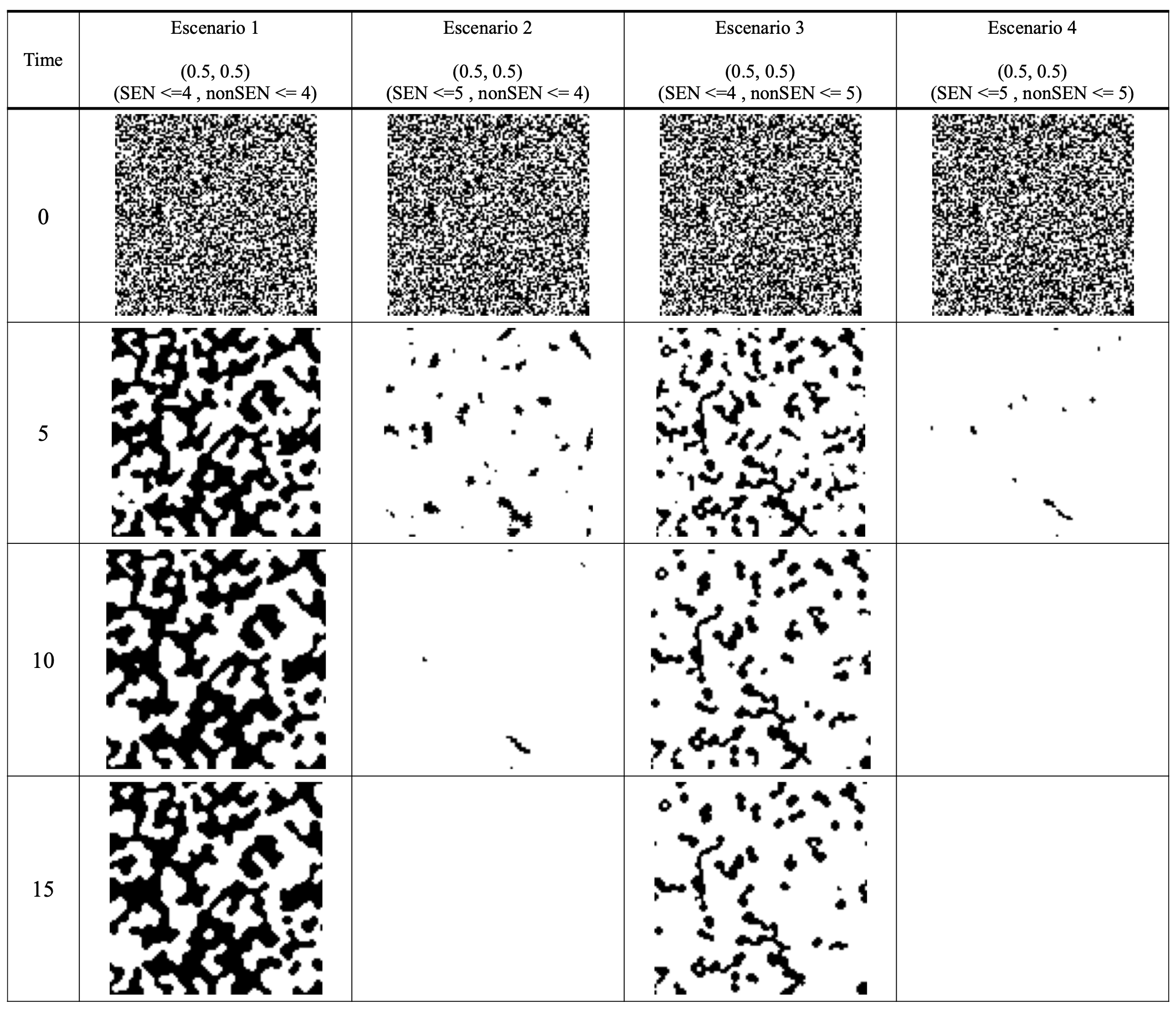}
	{\caption{Escenarios of empathy propagation. Escenario 1, segregation: $SEN_{students} = 4 $ and $nonSEN_{students} = 4$. Escenario 2: $SEN_{students} = 5 $ and $nonSEN_{students} = 4$. Escenario 3: $SEN_{students} = 4 $ and $nonSEN_{students} = 5$. Escenario 4: $SEN_{students} = 5 $ and $nonSEN_{students} = 5$.
	\label{es02}}}
\end{figure}

Data obtained in previous studies, in particular the work of \citet{RogersandMcKane2011}, identified different variations of the Shelling's model of segregation. In our study, we identify one of such variations with the case of inclusion in schools. In particular, we use the model of Shelling as our baseline scenario in which the segregation is the common pattern occurred in schools. Moreover, the model of cultural dissemination of \citet{Axelrod1984} was also used for showing the integration or inclusion between students with empathetic opinions. Therefore, our first result shows forth escenarios organized from the segregation to inclusive patterns (Figures \ref{es01} and \ref{es02}).

Figure \ref{es01} shows the density of empathy between students based on the percentage of students with the $nonSEN$ opinion. In particular, escenario 1 displays the classical segregation patterns related to the model of Shelling. Escenario 2 shows the case of empathy and inclusivity based on a direct influence of $SEN$ students. Compared with the escenario 1, this result shows a pattern behavior that describes a dissemination of inclusiveness. $SEN$ students show a direct and a higher peer pressure for accepting and changing opinions, then the percentage of $nonSEN$ students decreased significantly. Escenario 3 displays the case of empathy and inclusivity based on an indirect influence of $SEN$ students. It shows a medium effect (between segregation and inclusion) of empathy dissemination due to the indirect effect related to be less susceptible to change opinion. There are higher restrictions for changing an opinion when a student is self-considered $SEN$, then the percentage of $nonSEN$ students decreased moderately. Escenario 4 shows the mix of escenario 2 and 3. It displays the case of empathy and inclusivity based on a direct and indirect influence of $SEN$ students. This result outlines a situation in which the opinion of being $SEN$ is more accepted and less susceptible to change by students. This scenario reinforces the process of thinking about others as equals showing a faster convergence to being empathetic, that is the percentage of $nonSEN$ students decreased significantly.

In Figure \ref{es02}, we can see the escenarios of Figure \ref{es01} in a grid structure. Escenario 1 and 3 display the dynamics of segregation and the case of empathy and inclusivity based on an indirect influence of $SEN$ students. Both cases show a convergence and stable patterns of students with at least 20\% of empathy. Escenario 2 and 4 present the grid cases in which the convergence of empathy is completed. After $Time= 10$ the dissemination of empathy is accomplished.

To complement Figures \ref{es01} and \ref{es02}, Table \ref{nullmodels} shows the conceptual model validation that involves assessing the reliability of our escenarios. This result displays that our deterministic escenarios are statistically different from random realizations based on different cases of the binomial distribution. 

\begin{table}[!h]
\centering
\begin{tabular}{|c|c|c|c|}
\hline
{\bf Escenarios} & {\bf p=0.3} &{\bf p=0.5} & {\bf p=0.7}\\
\hline
1		& (0.2532, 0.3440)	& (0.0645, 0.8121)	& (-0.3007, 0.2576) \\
2		& (0.0203, 0.9402)	& (-0.4324, 0.0943)	& (-0.0458, 0.8660) \\
3		& (0.0203, 0.9402)	& (-0.4432, 0.0855)	& (-0.0458, 0.8661) \\
4		& (-0.0253, 0.9256)	& (-0.4543, 0.0770)	& (-0.0988, 0.7156) \\
\hline
\end{tabular}
\caption{\label{nullmodels}{\bf Correlation between escenarios and null models}. Columns with $p=0.3$, $p=0.5$, and $p=0.7$ show the probability of success for selecting $SEN$ students. This table shows the Spearman correlation between escenarios and  our proposed null models based on the binomial distribution in a fixed number of independent Bernoulli trials. Each tuple shows the Spearman correlation coefficient and its p-value. This measure computes the value associated with each deterministic escenario with the median of $10,000$ realizations per null model of the number of $nonSEN$ students. The probability density of the binomial distribution is the following: $p(N)=\tbinom{n}{N} p^{N}(1-p)^{n-N}$ where $n=$number of trails, $p=$probability of success, and $N=$number of successes. The probability mass function of the Bernoulli distribution is the following: $f(k)=\{1-p \quad if \quad k=0, p \quad if \quad k=1 \}$ for k in $\{0,1\}$, $0<= p <= 1$.
}
\end{table}

%\begin{table}[!h]
%\centering
%\begin{tabular}{|c|c|c|c|}
%\hline
%{\bf Escenarios} & {\bf uniform} &{\bf normal} & {\bf exponential}\\
%\hline
%1		& ({\bf 0.5208, 0.0385})	& (-0.0015, 0.9953)	& (0.1626, 0.5473) \\
%2		& (0.0851, 0.7538)	& ({\bf -0.5313, 0.0341})	& (0.1529, 0.5717) \\
%3		& (0.0851, 0.7539)	& ({\bf -0.5309, 0.0343})	& (0.1528, 0.5720) \\
%4		& (0.1407, 0.6031)	& ({\bf -0.5161, 0.0406})	& (0.1105, 0.6835) \\
%\hline
%\end{tabular}
%\caption{\label{nullmodels}{\bf Correlation between escenarios and null models}. This table shows the Spearman correlation between escenarios and  our proposed null models. Each tuple shows the Spearman correlation coefficient and its p-value. This measure computes the value associated with each deterministic escenario with the median of $10,000$ realizations per null model. The process of the normal and exponential models for generating 0's and 1's randomly is related to certain threshold in which if a value is less than a such threshold, return 1, otherwise 0. 
%The probability density function (PDF) of the uniform distribution is based on the following expression: $p(x)=1/b-a$ within the interval $[a=0, b=1)$. The PDF of the normal distribution is based on the following function: $p(x)= (1/ \sqrt{2 \pi \sigma^{2}}) e^{-(x-\mu)/ 2 \sigma^{2}}$ 
%$p(x)=\frac{1}{\sqrt{2 \phi \sigma^{2} }}e^{-\frac{(x-\mu)^{2}}{2 \sigma^{2}}}$
%where $\mu = 0$ and $\sigma^{2} = 1$. The PDF of the exponential distribution is based on the following function: $f(x) = exp(-x)$, for $x >= 1$.}
%\end{table}

We can see in Table \ref{nullmodels} positive and weak correlations in the fourth Escenarios when $p=0.3$. In the case when $p=0.5$, we see in Escenario 2, 3, and 4 negative and moderate correlations, meanwhile Escenario 1 displays a positive and a weak correlation. The last column shows the case when $p=0.7$, and it displays negative and weak correlations in the fourth Escenarios. Therefore, the four Escenarios are statically different from these null models.

% Furthermore, we see negative correlation between Escenarios 2, 3, and 4 with the normal distribution. The other cases display small correlation values indicating no monotonic relationships. These results indicate that our deterministic rules generate patterns different from random expectation.

\begin{figure}[!h]
	\centering
	\includegraphics[width=0.85\textwidth]{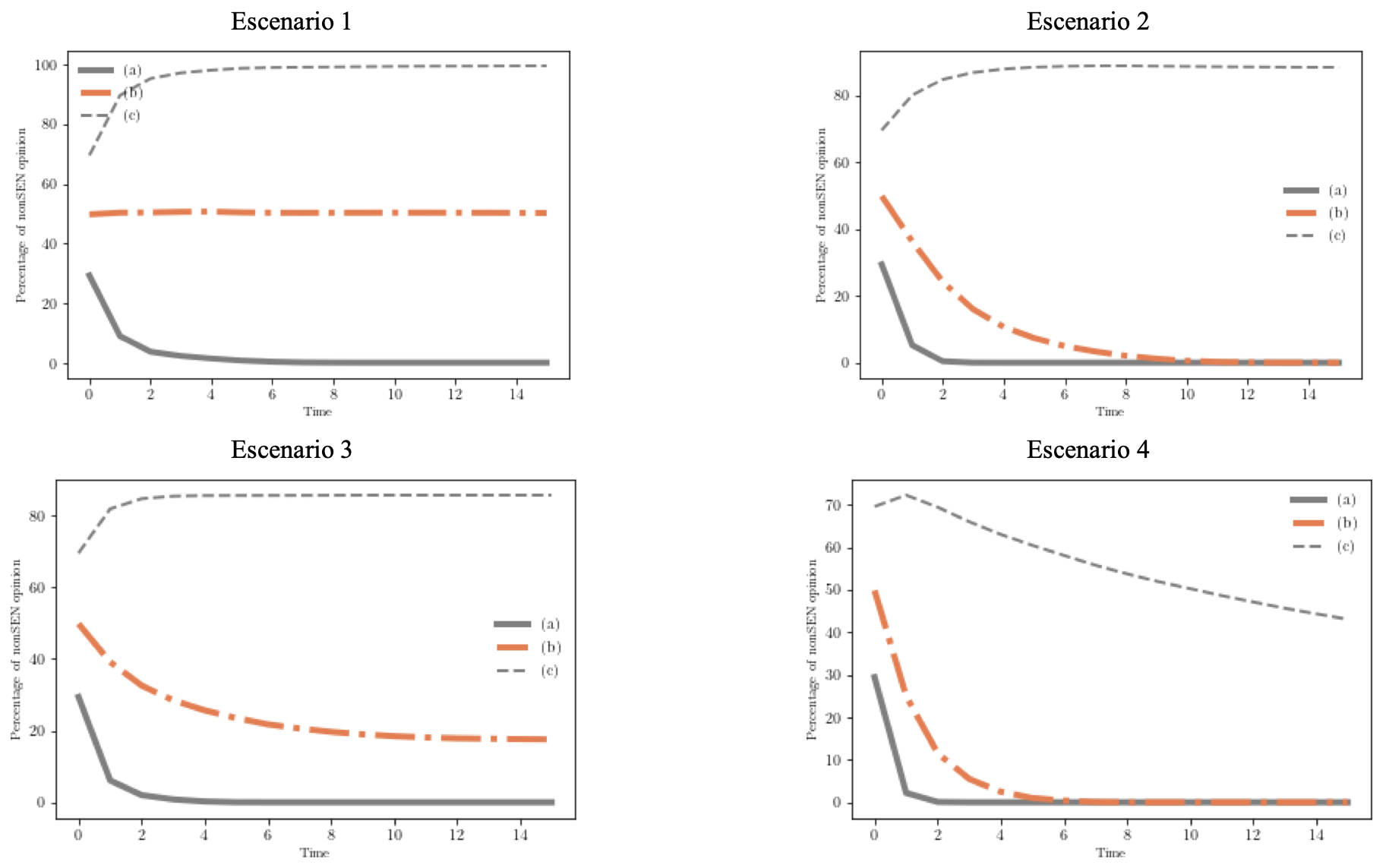}
	{\caption{Escenarios of empathy propagation with different proportions of students with and without $SEN$. (a) Proportion of 70\% and 30\% of students with and without $SEN$ respectively. (b) Base case, proportion equal to 50\% in both cases of students. (c) Proportion of 30\% and 70\% of students with and without $SEN$ respectively.
	\label{es03}}}
\end{figure}

Next, Figure \ref{es03} shows each escenario with different initial conditions. In particular, such variations are related to the proportion of students with and without $SEN$. In particular, variation (b) shows the base line related to Figures \ref{es01} and \ref{es02}. Variations (a) and (c) display the variants of setting different proportions of students with and without $SEN$.

We can see in Figure \ref{es03}, Escenario 1, different and stable patterns of segregation and empathy. The line associated with the variation (a) shows how a small proportion of $nonSEN$ students accelerates the process of change towards being self-perceived as $SEN$ student. The variation (b) displays the common result of a segregation pattern. The variation (c) shows how a higher proportion of $nonSEN$ students speeds up the process of reinforce their self-perception. 
Escenario 2, variations (a) and (b) show stable patterns of empathy. Both cases display an accelerated process of change towards being self-perceived as $SEN$ student. Variation (c) displays patterns of reinforce their self-perception as $nonSEN$ students. Escenario 3, variations (a) and (c) show similar patterns as Escenario 2. However, variation (b), which is the base line, clearly show a marked decrease in the percentage of self-perceiving as $nonSEN$ students. In addition, this variation shows a stable process of change towards being self-perceived as $SEN$ student that reach a low point around 20\%. 
Escenario 4 shows the best-case escenario in which the three variations reveal an empathy process related to the change towards being self-perceived as equal to others, $SEN$. Variations (a) and (b) show a marked decrease in the percentage of the self-perception as $nonSEN$. Variation (c) displays an interesting case that, at the beginning of the process, the percentage of the self-perception as $nonSEN$ increases, but subsequently it declines steadily.

Finally, Table \ref{nullmodels01} shows the statistical validation of the above escenarios and their variations (Figure \ref{es03}). In general, Table \ref{nullmodels01} displays from weak to very weak correlations. It suggests that our deterministic rules that generate patterns of inclusivity are statistical different from the null models.
In addition, we can see a negative direction in most of the correlation values. Almost 80\% of such values shows an inverse relationship.

\begin{table}[!h]
\centering
\begin{tabular}{|l|c|c|c|}
\hline
{\bf Escenarios} &  & & \\
{\bf and } & {\bf p=0.3} & {\bf p=0.5} & {\bf p=0.7} \\
{\bf variations} &  & & \\
\hline
1		& 	& 	&  \\
\hspace{1cm}(a)		& (-0.4707, 0.0656)	& (-0.1629,0.5464)	& (-0.2337, 0.3836) \\
\hspace{1cm}(c)		& (-0.0323, 0.9054)	& (-0.0767, 0.7774)	& (0.3617, 0.1685) \\
\hline
2		& 	& 	&  \\
\hspace{1cm}(a)		& (-0.3624, 0.1677)	& (-0.2820, 0.2899)	& (-0.5515, 0.0267) \\
\hspace{1cm}(c)		& (-0.2154, 0.4228)	& (-0.0477, 0.8605)	& (0.1894, 0.4822) \\
\hline
3		& 	& 	&  \\
\hspace{1cm}(a)		& (-0.4189, 0.1062)	& (-0.1009, 0.7099)	& (-0.2967, 0.2643) \\
\hspace{1cm}(c)		& (-0.0829, 0.7599)	& (-0.0131, 0.9614)	& (0.3828, 0.1432) \\
\hline
4		& 	& 	&  \\
\hspace{1cm}(a)		& (-0.2557, 0.3391)	& (-0.4866, 0.0559)	& (-0.5513, 0.0268) \\
\hspace{1cm}(c)		& (-0.0179, 0.9473)	& (0.0307, 0.9100)	& (-0.3293, 0.2128) \\
\hline
\end{tabular}
\caption{\label{nullmodels01}{\bf Correlation between variations of each escenario and null models}. (a) Proportion of 70\% and 30\% of students with and without $SEN$ respectively. (c) Proportion of 30\% and 70\% of students with and without $SEN$ respectively. Columns $p=0.3$, $p=0.5$, and $p=0.7$ show the probability of success or selecting $SEN$ students. 
Each tuple shows the Spearman correlation coefficient and its p-value between variations of each escenario and our proposed null models based on the binomial distribution in a fixed number of independent Bernoulli trials. The Spearman correlation computes the value associated with each variation of deterministic escenarios with the median of $10,000$ realizations per null model of the number of $nonSEN$ students.
}
\end{table}

These results therefore indicate that patterns of segregation of $SEN$ students in schools can change when there are small variations in local rules of interaction. In particular, the mix of a slightly higher peer pressure of $SEN$ students for accepting and changing opinions and a marginally higher restriction to be influenced from the $nonSEN$ students generates escenarios of inclusion via empathy.

\section{Discussion}
Nowadays, segregation patterns have still been shown in different social contexts. In particular, students with $SEN$ have commonly segregated and discriminated by surrounding people who supposed to be caring for them. 
Even thought, the inclusive education is on trend, it is not clear how small changes or variations in everyday behavior and thoughts can generate large-scale effects for the benefit of students with SEN. 
Therefore, in this theoretical study, we presented how a slightly variation in our self-perception of being $SEN$ can reverse the trend of segregation in schools.

In this study, we analyzed the possibility of increasing the empathic behavior between students in schools by using a virtual laboratory in which CAs generate digital models to conduct social experiments. This type of modeling is not new, but it still provides a better understanding of complex behaviors behind local influences and their emergent large-scale patterns. In particular, we can theoretically see that patterns of inclusivity via empathy between students in schools are possible if we are more empathic with $SEN$ students. Therefore, based on our research questions stated in the Introduction section, we can identify that one of the three deterministic rules can generate the best possible escenario. Escenario 4 is the desirable scenario because it shows how a small variation in the peer pressure of $SEN$ students and their restrictions to be influenced by the $nonSEN$ students generate inclusion patterns via empathy between students. In addition, the higher number of $SEN$ students at the initial conditions determines better results associated with inclusive patterns. 

Based on Figures \ref{es01}, \ref{es02}, and \ref{es03}, we can see that an empathy dissemination is more probable than ever before thanks to the current trend in inclusive education. We must remember that segregation is a learned behavior resulting from the tendency of people to differentiate themself to others based on social, cultural, or economical situations. Therefore, interpersonal interactions can reinforce our segregation patterns---conflicts---or our inclusive patterns---cooperation.

%To promote learning for all students, governments identify barriers to learning and participation associated with disabilities, among other learning, behavioral, and/or communication difficulties. In this way, governments point out that exclusion is not a problem for the students but of the schools. Therefore, differences in student abilities should not represent a barrier. These barriers are defined according to the school, classroom, family, and community contexts and can be attitudinal, pedagogical, or organizational. Attitudinal barriers are those related to segregation, exclusion, or attitudes of rejection or overprotection on the part of those who interact with the student (Gobierno del Estado de México, 2019).

An implication of these findings is the possibility to suggest local recommendations. Such recommendations should address issues for particular students, families, and educators communities. One of the most important aspects for suggesting local actions are the constant monitoring and evaluation of student with inclusive needs, for example feels respected, valued, and able to participate fully. In addition, we believe that an active participation of the private sector with local communities is fundamental to communicate efficiently such local initiatives. The following list describes our local recommendations:

\subparagraph{Local recommendations}
\begin{enumerate}
	\item This first recommendation is related to the escenario of empathy and inclusivity based on a direct influence of $SEN$ students. 
	\begin{enumerate}
	\item School board meetings should present information about inclusive education. 
	\item Every six months, school boards should update the information about inclusive education, for example new suggestions of international organizations and agencies, as well as information about programs implemented by the government.
	\item School boards should conduct and communicate the scientific evidence of successful cases of inclusive education. 
	\item School boards should coordinate educational campaigns to raise awareness about inclusive education via empathy between students.
	\item Schools should spend on training educators every six months specially those educators with limited empathy and tolerance. 
	\end{enumerate}
	
	\item This second recommendation is related to the escenario of empathy and inclusivity based on an indirect influence of $SEN$ students. 
	\begin{enumerate}
	\item Parent organizations should coordinate meetings for presenting information about inclusive education. 
	\item Parent organizations should be aware of local problems, for example the social instability, that can affect the interaction and learning conditions between students. 
	\item Parents should reinforce the self-perception of being $SEN$ or empathic by their example.
	\end{enumerate}
	
	\item This third recommendation is related to the escenario of empathy and inclusivity based on a direct and indirect influence of $SEN$ students.
	\begin{enumerate}
	\item Building a coalition with families and educators who support inclusive learning. 
	\item To a specific period of time, families and educators should share information about students' performance and the social context in the classroom. 
	\item Every three months, families and educators should monitor if the goals of inclusive education have been successfully achieved. 
	\item Every month, families and educators should create a questionnaire in current digital platforms, such as WhatsApp or third-party tools, for surveying and measuring student, families, and educators perceptions of inclusive behaviors. 
	\item Private sectors should support and promote awareness campaigns of inclusive behaviors based on TV, radio and digital media.
	\end{enumerate}
\end{enumerate}

%The government budget for people with disabilities should be addressed according to the needs of each age group.

%Budgetary reorganization is necessary, through a segmentation of resources that specifically includes spending on inclusive education, from educational programs and policies and with a vision from the Care System as indicated in Transversal Annex 31 of the Federal Expenditure Budget in Mexico.

In addition of these local recommendation, we suggest the following public policies or legislative proposals:
\subparagraph{Recommendation of public policy}
\begin{enumerate}
	\item The government has to spend a large amount of its budget in the education of children, in particular to the inclusive education. For example, the government should spend between 7.5\% and 9.5\% of its GDP in education.
	\item To effective allocate this funds, the government should prioritize high-impact areas as the inclusive education, for example educational programs related to family support, teacher training, and high-tech equipment.
	\item The government should design a more effective inclusive training program for educators in schools. In particular, creating inclusive learning environments for all students who can participate and feel a sense of belonging.
	
\end{enumerate}

Some limitations are worth noting. The gap between the educational policies and the local issues of schools will be a recurrent theme in the near future. 
Even thought our theoretical analysis showed a systematic review of local rules for generating stable patterns of inclusivity in schools, such rules can be related to successful cases only if there is a real coordination between students, families, educators, public and private sectors. Furthermore, the government should provide the legal and the best allocation of funds for supporting those local rules. Therefore, 
it is not enough to be politically correct in inclusive issues in schools; it is fundamental to apply real action plans with concrete steps for generating inclusive patterns in schools.

Future work should therefore include the design of public policies that permeates the whole educational system, in particular the inclusive education. For example, the government should have to generate the best action plan for spending a large amount of its budget in the education of children. It is highly probable that investing in early childhood generates long-term economic and social returns. Early interventions can generate rates of return higher than many investments in infrastructure or education at later stages. Therefore, investing in children is not an expense but an essential strategy for sustainable growth and social health.

%Investing in cognitive, socio-emotional, and physical development during the first years of life lays the foundation for a significant reduction in intergenerational inequalities.
%Educational exclusion is not an accidental phenomenon; it reflects a failure in the design, implementation, and funding of public policies, as well as institutional practices that perpetuate inequalities inside schools. 
\subsection{Conclusions}
Changing the segregation patterns to an inclusion behavior in schools is possible via the empathy between students. 
The mix of a slightly higher peer pressure---taking into account one more opinion of a student---in schools and a marginally higher restriction to be influenced from other students---endurance sense of self-perceived as being similar to others---can generate escenarios of inclusion via empathy.

%\section*{Submission date: November 20, 2025} 
%Scientific journal: International Journal of Inclusive Education\\ \href{https://www.tandfonline.com/toc/tied20/current}{https://www.tandfonline.com/toc/tied20/current}

%\section*{Fair Use Disclaimer}
%This document is for educational purposes only. Copyright Disclaimer under Section 107 of the Copyright Act of 1976: Allowance is made for ``fair use'' for purposes such as criticism, comment, news reporting, teaching, scholarship, education, and research. Fair use is a use permitted by copyright statute that might otherwise be infringing. All rights and credit go directly to its rightful owners. No copyright infringement is intended.

\section*{Disclosure statement}
The authors report there are no competing interests to declare.

\section*{Funding}
No funding was received for conducting this study.

\section*{Ethics approval}
This study did not require the local institutional review board (IRB) approval because this study does not include human research participants.

\section*{Consent to participate}
Informed consent for human research participants to be published in this study was not obtained because this study does not involve human participants.

\section*{Data availability}
The data is exclusive for scientific purposes. The data and the code that support the findings of this study are openly available in ``Empathy propagation and students with special education needs'' at \href{https://osf.io/z76kr/overview?view_only=fc75bc745f9e44e9a83a5b2292075300}{the Open Science Framework (OSF)}

%\section{References}

\end{document}